\documentclass[aps, showpacs, showkeys,nofootinbib,floatfix]{revtex4}

\usepackage{amssymb}
\usepackage{amsmath}
\usepackage{graphicx}
\usepackage[dvipdfm,colorlinks,citecolor=blue,urlcolor=blue]{hyperref}

\begin{document}

\title{\textbf{The Extended Analysis On New Generalized Chaplygin Gas }}\footnote{\small{Supported by the National Natural Science
Foundation of China (Grant No.10875056), and the Scientific Research
Foundation of the Higher Education Institute of Liaoning Province,
China (Grant No.2007T087 and 2009R35).}}

\author{WANG Jun\footnote{E-mail:wangjun\_3@126.com}, WU Ya-Bo\footnote{E-mail:
ybwu61@163.com}, WANG Di, YANG Wei-Qiang }

\affiliation{Department of Physics, Liaoning Normal University, Dalian 116029}

\begin{abstract}
We will extend the study of the new generalized Chaplygin gas (NGCG)
 based on [JCAP 0601(2006)003]. Concretely, we will not only discuss the change rates of the energy
densities and the energy transfer of this model, but also  perform
the $Om$ diagnostic to differentiate the $\Lambda$CDM model from the
NGCG and the GCG models. Furthermore, in order to consider the
influence of dark energy on the structure formation, we also present
the evolution of growth index in this scenario with interaction.

\end{abstract}


\maketitle

Recent observations of type Ia supernovae (SNe Ia)\cite{1,2}
indicate that the expansion of the Universe is accelerating at the
present time. From the observations of large scale structure
(LSS)\cite{3}, we know that the Universe is spatially flat. The
Wilkinson Microwave Anisotropy Probe (WMAP) indicates that our
Universe components are as follows: usual baryon matter occupies
about $4\%$, dark matter occupies about $23\%$ and dark energy
occupies about $73\%$. It is clear that the Universe is dominated by
an exotic component with large negative pressure, referred to as
dark energy. So many scientists believe the accelerating expansion
of the Universe is due to the dark energy.

~~Based on this opinion many dark energy models have been proposed.
The simplest form of dark energy is cosmological constant $\Lambda$
which would encounter "fine-tuning" problem and "coincidence"
problem. Other valid dark energy models are provided by scalar
fields, such as: Quintessence\cite{4,5}, which is introduced to
solve the "coincidence" problem and characterized by the equation of
state (EOS) $w_{de}$ between -1 and -1/3 (namely, $-1<w_{de}<-1/3$);
Phantom (ghost) field\cite{6}, which owns a negative kinetic
energy and characterized by the EOS $w_{de}$ less than -1 (namely,
$w_{de}<-1$); Tachyon field\cite{7,8}which can  act as a source of
dark energy depending upon the form of the tachyon potential, and so
on. Other scenarios on dark energy include brane world\cite{9},
generalized Chaplygin gas\cite{10}, holographic dark
energy\cite{11}, ect.

~~In the generalized Chaplygin gas (GCG) approach, dark energy and
dark matter can be unified by using an exotic equation of state.
This point can be easily seen from the fact that the GCG model
behaves as a dust-like matter at early times and as a cosmological
constant at late stage. Since the EOS of dark energy still can't be
determined exactly, the observational data show the EOS of dark
energy is in the range of ($-1.38,-0.86$)\cite{12,13}, the GCG
model is generalized to accommodate any possible X-type dark energy
with constant $w_{de}$. The new generalized Chaplygin gas (NGCG) as
a dark energy model is proposed for unification of X-type dark
energy and dark matter\cite{14}. The EOS of this model is
$p=-\tilde{A}(a)/\rho^{\alpha}$, where $a$ is the scale factor and
$\tilde{A}(a)=-w_{de}Aa^{-3(1+w_{de})(1+\alpha)}$. We know that the
NGCG model behaves as a dust-like matter at early times and as a
cosmological constant at late stage. Also this model is a kind of
the interacting XCDM model\cite{14}. Based on the Ref.\cite{14},
in this paper we will extend the analysis on the NGCG model by
discussing the change rates of the energy densities for dark energy
(DE) and dark matter (DM) as well as the energy transfer. Moreover,
we perform the $Om$ diagnostic to differentiate the $\Lambda$CDM
model from the NGCG and the GCG models. Furthermore, in order to
consider the influence of DE on the structure formation in this
scenario, we present the evolution of growth index in this model
with interaction.

In the framework of FRW cosmology, the EOS of the NGCG model is
described as\cite{14}
\begin{equation}\label{1}
p_{NGCG}=-\frac{\tilde{A}(a)}{\rho^{\alpha}_{NGCG}},
\end{equation}
where $\alpha$ is a real number and $\tilde{A}(a)$ is a function
depending on the scale factor of the Universe. The form of the
function $\tilde{A}(a)$ is:
\begin{equation}\label{2}
\tilde{A}(a)=-w_{de}Aa^{-3(1+w_{de})(1+\alpha)},
\end{equation}
where A is a positive constant and $w_{de}$, which should be taken
as any possible value in the range (-1.38, -0.86), is a constant. We
can see explicitly that when $w_{de}=-1$, the EOS can reduce to the
GCG scenario, while when $\alpha=0$ the XCDM model can be obtained
again.

In the following discussions  we suppose that in the NGCG model the
fluid is made up of two components, one is the DE component marked
as $\rho_{de}$, and the other is the DM component marked as
$\rho_{dm}$, i.e.
\begin{equation}\label{3}
\rho_{NGCG}=\rho_{de}+\rho_{dm},
\end{equation}
\begin{equation}\label{4}
p_{NGCG}=p_{de}.
\end{equation}

Then, the energy density can be expressed as\cite{14}
\begin{equation}\label{5}
\rho_{NGCG}=\rho_{NGCG0}a^{-3}[1-A_{s}+A_{s}a^{-3w_{de}(1+\alpha)}]^{\frac{1}{1+\alpha}},
\end{equation}
where $A_{s}=\frac{\Omega^{0}_{de}}{1-\Omega^{0}_{b}}$. The energy
densities of the DE and the DM components can be respectively
expressed as
\begin{equation}\label{6}
\rho_{de}=\rho_{de0}a^{-3[1+w_{de}(1+\alpha)]}[1-A_{s}+A_{s}a^{-3w_{de}(1+\alpha)}]^{\frac{1}{1+\alpha}-1},
\end{equation}
\begin{equation}\label{7}
\rho_{dm}=\rho_{dm0}a^{-3}[1-A_{s}+A_{s}a^{-3w_{de}(1+\alpha)}]^{\frac{1}{1+\alpha}-1}.
\end{equation}

Below we study the change rates of the energy densities of DE and
DM. First, we consider the case without interaction between them. By
means of the continuity equation, the energy densities of DM and DE
can be respectively obtained as follows:
\begin{equation}\label{8}
\rho_{dm}=\rho_{dm0}(1+z)^{3} ,
\end{equation}
\begin{equation}\label{9}
\rho_{de}=\{\rho_{NGCG0}[1-A_{s}+A_{s}(1+z)^{3w_{de}(1+\alpha)}]^{\frac{1}{1+\alpha}}-\rho_{dm0}\}(1+z)^{3}
.
\end{equation}
Thus, the change rates of the energy densities of DE and DM can be
obtained:
\begin{equation}\label{10}
\frac{d\rho_{de}}{dz}=3\rho_{NGCG0}M ,
\end{equation}
\begin{equation}\label{11}
\frac{d\rho_{dm}}{dz}=3\rho_{dm0}(1+z)^{2},
\end{equation}
where
$M=(1+z)^{2}[1-A_{s}+A_{s}(1+z)^{3w_{de}(1+\alpha)}]^{\frac{1}{1+\alpha}}(1+\frac{w_{de}A_{s}(1+z)^{3w_{de}(1+\alpha)}}{1-A_{s}+A_{s}(1+z)^{3w_{de}(1+\alpha)}})$
and $A_{s}=\frac{\Omega^{0}_{de}}{1-\Omega^{0}_{b}}=0.74$ (the
current parameters used in this paper are: $\Omega_{dm0}=0.25$,
$\Omega_{de0}=0.7$, and $\Omega_{b0}=0.05.)$. It is easy to see that
the sign of the value of $\frac{d\rho_{de}}{dz}$ lies on M. Hence
the evolutional trend of DE density can be discussed by the factor M
as well.

\begin{figure}[!ht]
\begin{center}
\includegraphics{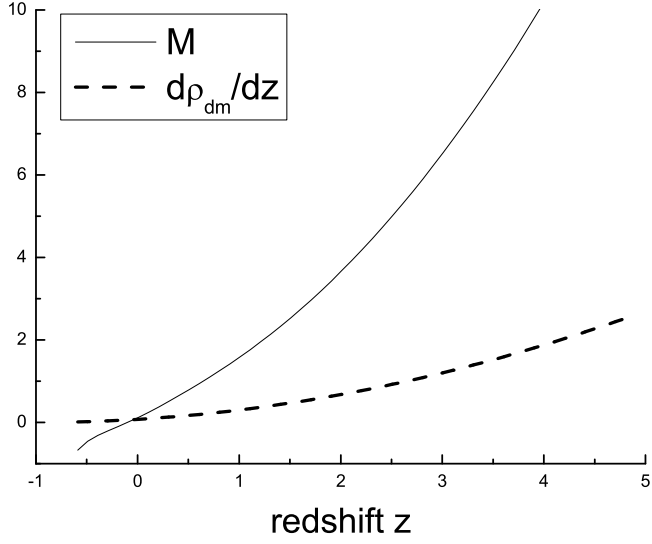}
\end{center}
\caption{The evolutional trends of the factor M and
$\frac{d\rho_{dm}}{dz}$ with $z$. Here we choose $A_{s}=0.74$,
$\alpha=0.5$, $w_{de}=-1.2$.}
\label{F1}
\end{figure}

The evolutional trends of the factor M and $\frac{d\rho_{dm}}{dz}$
are illustrated in Fig.\ref{F1}, from which we can see that when
$z>0$, the change rate of the energy density of DE decreases faster
than that of DM as the redshift $z$ becomes low. When $z<0$, it
keeps decreasing but $\frac{d\rho_{dm}}{dz}$ would approach to $0$.

\begin{figure}[!ht]
\begin{center}
\includegraphics{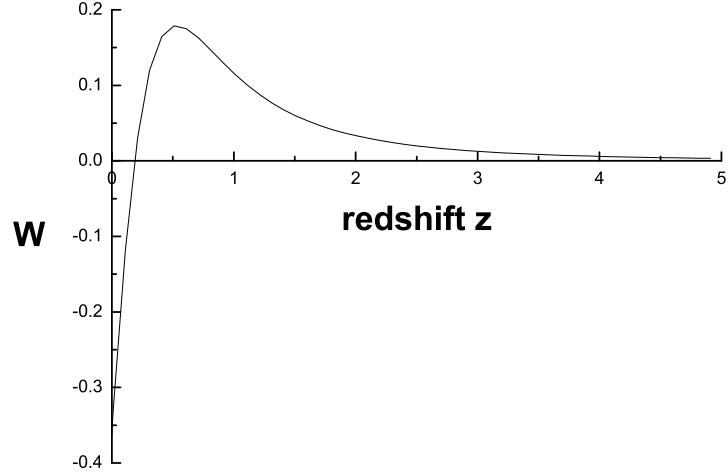}
\end{center}\caption{the relation between factor W and the redshift $z$. Here we choose $A_{s}=0.74$,
$\alpha=0.5$, $w_{de}=-1.2$.}
\label{F2}
\end{figure}

In what follows we will consider the NGCG model with interaction
between DE with constant $w_{de}$ and DM. According to the
Eqs.(\ref{6}) and (\ref{7}), we can give the expressions of the
change rates of the energy densities of DE and DM as follows:
\begin{equation}\label{12}
\frac{d\rho_{de}}{dz}=3\rho_{de0}W,
\end{equation}
\begin{equation}\label{13}
\frac{d\rho_{dm}}{dz}=3\rho_{dm0}S,
\end{equation}
where
$W=(1+z)^{3w_{de}(1+\alpha)+2}[1-A_{s}+A_{s}(1+z)^{3w_{de}(1+\alpha)}]^{\frac{1}{1+\alpha}-1}[1+w_{de}(1+\alpha)-\frac{\alpha
w_{de}A_{s}}{1-A_{s}+A_{s}(1+z)^{3w_{de}(1+\alpha)}}]$,
\\
$S=(1+z)^{2}[1-A_{s}+A_{s}(1+z)^{3w_{de}(1+\alpha)}]^{\frac{1}{1+\alpha}-1}
(1-\frac{\alpha
w_{de}A_{s}(1+z)^{3w_{de}(1+\alpha)}}{1-A_{s}+A_{s}(1+z)^{3w_{de}(1+\alpha)}})$.
From Eq.(\ref{12}), it is easy to see that the sign of the value of
$\frac{d\rho_{de}}{dz}$ depends on the factor W. Fig.\ref{F2} shows
the relation between factor W and the redshift $z$.

From Fig.\ref{F2}, it is easy to see that as the redshift $z$
becomes low the change rate of the energy density of DE is from
increasing gradually to decreasing sharply. Specially, when $z > 4$
it would approach to $0$. It arrives at the maximum when $z=0.53$.
While when $z=0.2$ it vanishes. It follows that the point of
$z=0.53$ is a transformation point. Note that there is not any
transformation point in the case without interaction.

\begin{figure}[!ht]
\begin{center}
\includegraphics{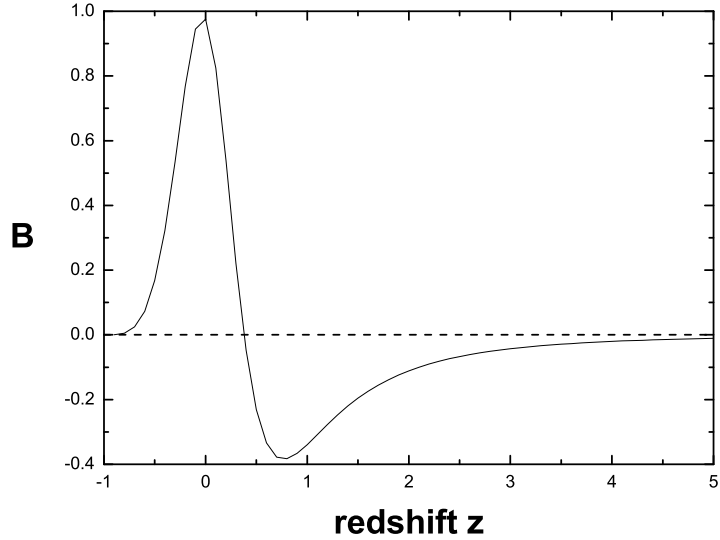}
\end{center}\caption{The
relation between factor B and the redshift $z$. Here we choose
$A_{s}=0.74$, $\alpha=0.5$, $w_{de}=-1.2$.}
\label{F3}
\end{figure}

From the above discussions, we see explicitly that there would exist
energy transfer between DE and DM. Now, we consider a feasible
interaction with an ansatz $\Gamma=3Hc^{2}\rho$ \cite{15}, where
$\rho$ is the total energy density and $c^{2}$ is a constant
denoting the transfer strength. We suppose the components of the
NGCG interact through the interaction term $\Gamma$. The components
of the NGCG respectively satisfy
\begin{equation}\label{14}
\dot{\rho}_{de} +3H(1+w_{de})\rho_{de}=-\Gamma,
\end{equation}
\begin{equation}\label{15}
\dot{\rho}_{dm}+3H\rho_{dm}=\Gamma.
\end{equation}
Note that $\Gamma>0$, which implies there is an energy transfer from
DE to DM. By using Eqs.(\ref{7}), (\ref{13}) and (\ref{15}) as well
as $\frac{dz}{dt}=-(1+z)H$,  we can obtain:
\begin{equation}\label{16}
\dot{\Gamma}=-9H\rho_{dm0}A_{s}w_{de}\alpha B,
\end{equation}
where
$B=(1+z)^{3w_{de}(1+\alpha)+2}[1-A_{s}+A_{s}(1+z)^{3w_{de}(1+\alpha)}]^{\frac{1}{1+\alpha}-2}[1+w_{de}(1+\alpha)-
\frac{(1+2\alpha)A_{s}w_{de}(1+z)^{3w_{de}(1+\alpha)}}{1-A_{s}+A_{s}(1+z)^{3w_{de}(1+\alpha)}}]
$. From Eq.(\ref{16}), we can see that the evolution of
$\dot{\Gamma}$ can be described by the factor B. The relation
between factor B and the redshift $z$ is shown in Fig.\ref{F3}.

According to Fig.\ref{F3}, we can see that when $z>0.38$ the
transfer direction of the energy flow is from DM to DE and the
quantity of energy transfer is from increasing gradually to
decreasing sharply as the redshift $z$ becomes low. But when
$-1<z<0.38$ the transfer direction of the energy flow just reverses
and the quantity of energy transfer is from increasing to decreasing
sharply, until approaches to $0$ at $z=-1$. Clearly, the point of
$z=0.38$ is the transformation point and the quantity of energy
transfer doesn't change any more at this point. Specially, when
$z=0.78$, the quantity of energy transfer from DM to DE reaches to
the maximum. While when $z=-0.02$,  it reaches to the maximum from
DE to DM. It follows that the above discussions not only show the
evolutionary laws of the energy transfer in the past and at present,
but also predict the situation in the future.

Below, we will determine the present value of $c^{2}$. The ratio of
the energy density can be defined as
$\dot{r}=\frac{\rho_{dm}}{\rho_{de}}$. Thus by means of
Eqs.(\ref{6}),(\ref{7}),(\ref{14}) and (\ref{15}) the evolution
equation of ratio r can be expressed as
\begin{equation}\label{17}
\dot{r}=\frac{\rho_{dm}}{\rho_{de}}(\frac{\dot{\rho_{dm}}}{\rho_{dm}}-\frac{\dot{\rho_{de}}}{\rho_{de}})
=3Hr[c^{2}(\frac{1}{\Omega_{dm}}+\frac{1}{\Omega_{de}})+w_{de}],
\end{equation}
from which, the following expression is easily obtained:
\begin{equation}\label{18}
c^{2}=[\frac{-w_{de}(1+\alpha)}{(1+z)H}+w_{de}](\frac{1}{\Omega_{dm}}+\frac{1}{\Omega_{de}})^{-1}.
\end{equation}
Hence we can obtain the corresponding present values of $c^{2}$  to
the different values of $w_{de}$ and $\alpha$ shown in Tab.
\ref{T1}.
\begin{table}[!ht]
\centering{\small{\begin{tabular}{c c c c} \hline\hline
 $w_{de}$ & $-0.8$ & $-1$ & $-1.2$
\\ \hline $c^{2}$ & $-0.0193076$ & $-0.0242937$ & $-0.0292797$ \\
\hline \hline \multicolumn{3}{c}{(a)$\alpha=0.5$}
\end{tabular}
\begin{tabular}{c c c c} \hline\hline $\alpha$
&$0$&$0.2$ & $0.5$  \\ \hline $c^{2}$ & $-0.0242937$ &
$-0.0242937$ & $-0.0242937$ \\
\hline\hline\multicolumn{3}{c}{(b)$w_{de}=-1$}
\end{tabular}}}
\caption{\small{The corresponding present values of $c^{2}$. The
current parameters are taken to be $z=0$, $\Omega_{dm0}=0.25$,
$\Omega_{de0}=0.7$, $H_{0}=70.5$\cite{13}.}}\label{T1}
\end{table}

From Tab.\ref{T1}(b), it is easy to see that the values of $c^{2}$
don't change with $\alpha$ just as the case in the $\Lambda$CDM
model.

Moreover, we will use a new geometrical diagnostic method, $Om$
diagnostic\cite{20}, to differentiate the $\Lambda$CDM model from
the NGCG and the GCG models. The $Om$ diagnostic is defined as
follows
\begin{equation}\label{19}
Om(x)=\frac{E^{2}(x)-1}{x^{3}-1}, x=1+z,
\end{equation}
where
$E^{2}(x)=H(x)/H_{0}=(1-\Omega_{b0})x^{3}[1-A_{s}+A_{s}x^{3w_{de}(1+\alpha)}]^{\frac{1}{1+\alpha}}+\Omega_{b0}x^{3}$
for the NGCG model.

\begin{figure}[!ht]
\begin{center}
\includegraphics{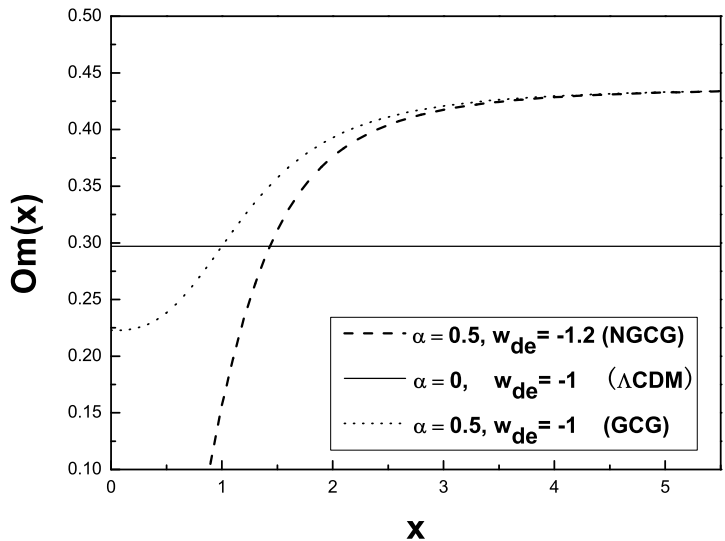}
\end{center}\caption{The $Om(x)$ evolution diagram.
The current parameters are taken to be $\Omega_{dm0}=0.25$,
$\Omega_{de0}=0.7$, $\Omega_{b0}=0.05$ and $A_{s}=0.74$.}
\label{F4}
\end{figure}

For the $\Lambda$CDM model, $Om(z)=\Omega_{om}$ is a constant. It
provides a null test of cosmology constant. The benefit of $Om$
diagnostic is that the quantity $Om(z)$ can distinguish DE models
with less dependence on matter density $\Omega_{m0}$ relative to the
EOS of DE\cite{20}. We can get the expression of $Om(x)$ in the
NGCG model as
\begin{equation}\label{20}
Om(x)=\frac{(1-\Omega_{b0})x^{3}[1-A_{s}+A_{s}x^{3w_{de}(1+\alpha)}]^{\frac{1}{1+\alpha}}+\Omega_{b0}x^{3}-1}{x^{3}-1}
\end{equation}
where $A_{s}=0.74$. It follows that when taking $\alpha=0.5,
w_{de}=-1$, Eq.(\ref{20}) can reduce into the $Om(x)$ of the GCG
model, but when taking $\alpha=0, w_{de}=-1$, it corresponds to the
$Om(x)$ of the $\Lambda$CDM model. For the NGCG model, the curve of
$Om(x)$ corresponding to $\alpha=0.5$, $w_{de}=-1.2$ is plotted in
Fig.\ref{F4}.

The Fig.\ref{F4} shows that when $x\geq4$ the evolutionary
trajectory of the NGCG model is similar to one of the GCG model, but
in the recent period the both become very different. It follows that
the both NGCG and GCG models are completely different from the
$\Lambda$CDM model.

Furthermore, in order to study the influence of the NGCG model on
the structure formation, now we discuss the growth index of the NGCG
model. The growth index is an important quantity to test a model,
which can be measured by the galaxy correlation function or the
peculiar velocities. Its definition is as follows:
\begin{equation}\label{21}
f\equiv\frac{d\ln\delta}{d\ln a}=\frac{a}{\delta}\frac{d\delta}{da},
\end{equation}
where $\delta\equiv\frac{\delta\rho_{m}}{\rho_{m}}$ is the matter
density contrast and $a$ is the scale factor of the Universe.

According to the perturbation equation\cite{16}, the growth index
$f$ satisfies the following equation:
\begin{equation}\label{22}
f'+\frac{f^{2}}{a}+(\frac{2}{a}-1)f-\frac{3}{2a}\Omega_{m}=0,
\end{equation}
where the prime denotes the derivation with respect to the scale
factor $a$. In the NGCG model, we take
$\Omega_{m}=\Omega_{dm}+\Omega_{b}$. the corresponding expression of
$\Omega_{dm}$ and $\Omega_{b}$ reads\cite{14}:
\begin{equation}\label{23}
\Omega_{dm}=\Omega_{dm0}E^{-2}a^{-3}[1-A_{s}+A_{s}a^{-3w_{de}(1+\alpha)}]^{\frac{1}{1+\alpha}-1},
\end{equation}
\begin{equation}\label{24}
\Omega_{b}=\Omega_{b0}E^{-2}a^{-3},
\end{equation}
where
$E=\{(1-\Omega_{b0})a^{-3}[1-A_{s}+A_{s}a^{-3w_{de}(1+\alpha)}]^{\frac{1}{1+\alpha}}+\Omega_{b0}a^{-3}\}^{1/2}$
and $A_{s}=0.74$.

\begin{figure}[!ht]
\begin{center}
\includegraphics{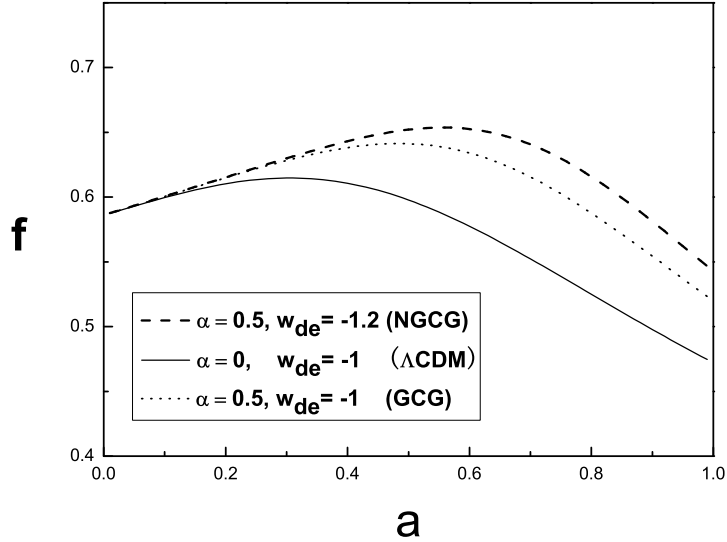}
\end{center}
\caption{The evolutionary trajectories of the growth indices $f$ for the NGCG, the GCG and the $\Lambda$CDM models.
The current parameters are taken to be $\Omega_{dm0}=0.25$,
$\Omega_{de0}=0.7$, $\Omega_{b0}=0.05$ and $A_{s}=0.74$.}
\label{F5}
\end{figure}

The evolutionary trajectory of the growth index $f$ for the NGCG
model is plotted in Fig.\ref{F5}. By calculations, the value of $f$
for the NGCG model is $0.592$ at $a=0.87$ (namely, $z=0.15$). From
the 2dF galaxy redshift survey (2dFGRS)\cite{17,18,19}, we know
that the observational value of $f$ is $f=0.51\pm0.1$ or
$f=0.58\pm0.11$ at the effective redshift of the survey $z = 0.15$.
It means the theoretical value of growth index given by us is
consistent with observation. In Fig.\ref{F5}, we also give the
evolutionary trajectories of growth indices of the GCG model
(namely, $\alpha=0.5, w_{de}=-1$) and the $\Lambda$CDM model
(namely, $\alpha=0, w_{de}=-1$). By comparing these three
evolutionary trajectories in Fig.\ref{F5}, we can find that at early
times, the growth indices of these three models are the same; but
near $a=0.09$, the evolutionary trajectory of $\Lambda$CDM model
would be different from the other two; while near $a=0.2$, the
evolutionary trajectory of the GCG model would deviate from one of
the NGCG model.

In summary, we have extended the analysis on the new generalized
Chaplygin gas (NGCG) model as the unification of DM and DE.
Concretely, we have studied the evolutional trends of the energy
densities, $Om$ diagnostic and the growth index in the NGCG model.
By complicated calculations and analysis, we can give some comments
as follows:

$1.$ In the NGCG model without interaction between DM and DE, the
change rate of the energy density of DE decreases gradually as the
redshift $z$ becomes low, but it would be from increasing gradually
to decreasing sharply in the case with interaction, while $z = 0.53$
is a transformation point.

$2.$ For the NGCG model with interaction term $\Gamma=3Hc^{2}\rho$,
the evolutionary laws of the energy transfer have been discussed.
Furthermore, we have also given the present values of $c^{2}$ when
taking a fixed constant $\alpha$ (or $w_{de}$), but different values
of $w_{de}$ (or $\alpha$). The results show that for a fixed
$w_{de}$, the present values of $c^{2}$ don't change with $\alpha$,
which is as the same as one of the $\Lambda$CDM model.

$3.$ We have performed the $Om$ diagnostic to the NGCG model and
shown the discrimination among the NGCG, the GCG and the
$\Lambda$CDM models.

$4.$ The evolutionary trajectory of the growth index $f$  has also
been illustrated for the NGCG model with interaction. We find the
value of $f$ at $a=0.87$ (i.e., $z=0.15$) given by us agrees with
the observational value of $f=0.51\pm0.1$ or $f=0.58\pm0.11$ at the
effective redshift of the survey $z=0.15$. By comparing the NGCG and
the GCG models with the $\Lambda$CDM model, we have shown that at
early times, the growth indices of these three models are the same.
Near $a =0.09$, the evolutionary trajectory of $\Lambda$CDM model
would be different from ones of the other two. Near $a=0.2$, the
evolutionary trajectory of GCG model would deviate from one of the
NGCG model.

\end{document}